\documentclass[conference]{IEEEtran}
\usepackage{color}
\usepackage{algorithm}
\usepackage{cite}
\usepackage{graphicx}
\usepackage{amsmath}
\usepackage{amssymb}
\usepackage{algorithmic}
\usepackage{tikz}
\usepackage{geometry}
\usepackage{hyperref}

\newcommand\copyrighttext{%
\footnotesize \textcopyright 2020 IEEE. Personal use of this material is permitted. Permission from IEEE must be obtained for all other uses, in any current or future media, including reprinting/republishing this material for advertising or promotional purposes, creating new collective works, for resale or redistribution to servers or lists, or reuse of any copyrighted component of this work in other works. DOI: \href{https://ieeexplore.ieee.org/document/9181119}{10.1109/ISCAS45731.2020.9181119}}
\newcommand\copyrightnotice{%
\begin{tikzpicture}[remember picture,overlay]
\node[anchor=south,yshift=10pt] at (current page.south) {\fbox{\parbox{\dimexpr\textwidth-\fboxsep-\fboxrule\relax}{\copyrighttext}}};
\end{tikzpicture}%
}

\begin{document}

\title{A Proposal for Multiplierless RLS Adaptive Filters with Implementation Complexity Constraint}

\author{\IEEEauthorblockN{Luiz Felipe da S. Coelho, Lisandro Lovisolo and Michel P. Tcheou}
\IEEEauthorblockA{Department of Electronics and Communications Engineering\\
Rio de Janeiro State University\\
20550-900 Rio de Janeiro, RJ, Brazil\\
Email: luizfelipe.coelho@smt.ufrj.br, lisandro@uerj.br and mtcheou@uerj.br}}

\newgeometry{top=.7in,bottom=0.9in,right=0.7in,left=.7in}
\maketitle
\copyrightnotice

\begin{abstract}
The so-called sum-of-powers-of-two arithmetic has the advantage of using simple shift and thus avoiding multiplications operations. In this work, we provide a feasible RLS adaptive filter algorithm employing the SOPOT representation. This reduces the computational burden of the RLS algorithm. Therefore, leading to a reduction in the power consumption from a hardware implementation point of view, which is an important feature in the developing of smaller, lighter and more power-efficient portable devices.
\end{abstract}

\IEEEpeerreviewmaketitle

\section{Introduction}

\textit{Sums-of-Powers-of-Two} (SOPOT) have been used to approximate the filter coefficients aiming for computational complexity reduction~\cite{samueli1989improved, evans1994efficient, macleod2005multiplierless, lim1999signed, yeung2002multiplier,da2014fir}. This results is power-saving that may increase battery life expectancy of devices and  sensors. In this work, one devises how to implement the \textit{Recursive Least Squares} (RLS) adaptive filter using uniquely SOPOT arithmetic, meaning that the coefficient representation and its update rely solely on SOPOT, providing a fully operational adaptive filter with SOPOT arithmetic. 

\begin{figure}[hbt!]
    \centering
    \includegraphics[width=.8\linewidth]{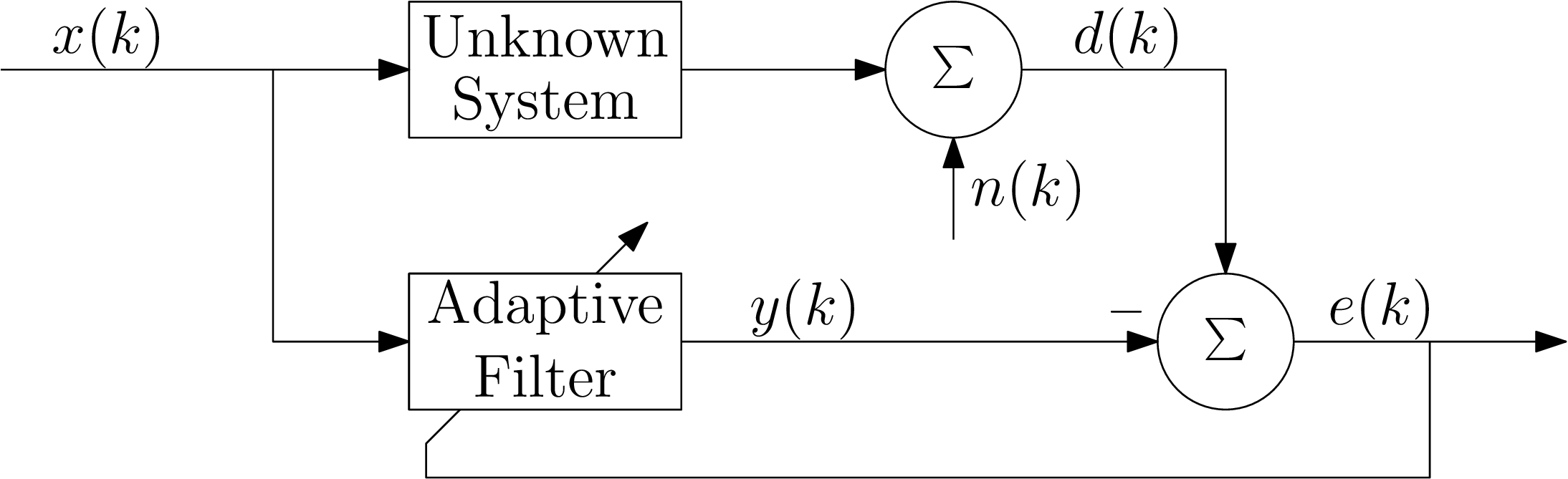}
    \caption{Adaptive filter in a system identification application.}
    \label{fig:adapt_filt}
\end{figure}

The use of SOPOT representation imposes some numerical restrictions inherited from its finite-precision. These may worsen as computational complexity restriction increases by reducing the amount of terms used in the numeric representation. Consequently, the effects of using SOPOT for implementing RLS adaptive filters are evaluated in a system identification application, as shown in Fig. \ref{fig:adapt_filt}.

This paper is organized by first bringing the RLS and the SOPOT arithmetic in Section~\ref{sec:proposed_spt-rls}.
In sequence, Section \ref{sec:implementing} provides the specificities for implementing the RLS algoritm using SOPOTs and Section~\ref{sec:nmerical} discusses specific numerical impairments that may arise from that. Subsequently, in Section~\ref{sec:results} we present some simulation results and in Section~\ref{sec:conclusion} the work conclusion is presented.

\section{SOPOT based RLS Adaptive Filter for System Identification} \label{sec:proposed_spt-rls}

\subsection{SOPOT Arithmetic}

In the SOPOT scenario, a value is approximated using a sum of \textit{Signed Power-of-Two} (SPT) terms.  Let $v$ be a scalar value it is approximated using $M$ SPT terms $v'$ by means of
\begin{align}
    v' =\sum_{m=0}^{M-1} v_{m}2^{b_{v_m}} = v_{0} 2^{b_{v_{0}}} + v_{1} 2^{b_{v_{1}}} +~...~ + v_{M-1} 2^{b_{v_{M-1}}} \label{eq:SOPOT_notation}
\end{align}
where $v_{m}$ and $b_{v_m}$ denote, respectively, the sign (+ or -) and the exponent of the $m$-th SPT term (an integer). One may assume that $B$ bits are used to encode each exponent and that $L$ is the largest exponent value, i.e., $L+1-2^B \leq b_{v_m} \leq L$. The smaller the number of terms $M$ employed, the greater approximation errors are expected. On the other hand, the larger is the reduction in complexity.

Different representations are possible for some values using, equation (\ref{eq:SOPOT_notation}) since ($2^p - 2^{p-1} = 2^{p-1}$). The interplay among SPT terms leads to nonuniform quantization as the incurred error may depend on the value to be represented~\cite{li1994multiplierless}. 
Consequently, algorithms for obtaining the approximation for a given number have been proposed~\cite{lim1991decomposition,li1994multiplierless,da2014fir}. For each SPT term, a single bit can be used for the sign while $B$ bits are used to store the exponent, therefore for a value represented using $M$ terms, $M(B+1)$ bits are required. 

\subsection{The RLS Algorithm}

Let $x(k)$ be the input signal for the adaptive filter, $d(k)$ the desired signal, $y(k)$ the output signal and $e(k)$ the error signal in the $k$-th iteration or signal sample. The adaptive filter of order $N-1$ (employing $N$ coefficients) can be represented as a vector of weights or weight vector ${\bf w}(k) = [w_0(k)~w_1(k)~...~w_{N-1}(k)]^T$. The RLS algorithm updates this iteratively by means of
\begin{align}
    {\bf S}_D(-1) &= \delta {\bf I}_{N} \textrm{ and } {\bf w}(-1) = [0~0~...~0]^T\\ 
    e(k) &= d(k) - {\bf x}^T(k){\bf w}(k-1) \label{eq:RLS_error} \\
    {\bf \Psi}(k) &= {\bf S}_D(k-1){\bf x}(k) \label{eq:RLS_psi} \\
    {\bf S}_D(k) &= \frac{1}{\lambda} \left[ {\bf S}_D(k-1) - \frac{ {\bf \Psi}(k) {\bf \Psi}^T(k)}{\lambda + {\bf \Psi}^T(k) {\bf x}(k)} \right] \label{eq:RLS_SD} \\
    {\bf w}(k) &= {\bf w}(k-1) + e(k){\bf S}_D(k){\bf x}(k), \label{eq:RLS_w}
\end{align}
where $\delta$ is the initialization value for the deterministic correlation matrix ${\bf S}_D$, $\lambda$ is the forgetting factor and the vector $\bf x$ is the tapped delay line of the input signal containing its last $N$ samples, i.e., ${\bf x}(k)~=~[x(k)~x(k-1)~...~x(k-N+1)]^T$.

Not only is the RLS adaptive filter known for its fast convergence but also for being computationally demanding~\cite{dinizadaptive}. For using it in small power-demanding devices, it may be interesting to reduce the computational burden. For the \textit{Signed Power-of-Two Recursive Least Squares} (SPT-RLS) adaptive filter, SOPOT arithmetic is used both to compute the output and the filter coefficients. This requires the computation of products between vectors, between matrix and vector, and between matrices.

\subsection{SOPOT Inner Products}

An inner product is the sum of element-wise multiplications of vectors. Therefore, this element-wise multiplication is defined as the key-operation in the inner and matrix products. Using the notation in equation (\ref{eq:SOPOT_notation}) one clearly sees that the multiplication of two SOPOT values $v' u'$ provides
\begin{align}
    v' u' = ~ & v_{0}u_{0}2^{b_{v_{0}}+b_{u_{0}}} +
    v_{0}u_{1}2^{b_{v_{0}}+b_{u_{1}}} + \cdots + \label{eq:multiply} \\
    +~&  v_{1}u_{0}2^{b_{v_{1}}+b_{u_{0}}} + \cdots + 
    v_{M-1}u_{M-1}2^{b_{v_{M-1}}+b_{u_{M-1}}} \notag
\end{align}
and the multiplication process turns into additions of SPT terms exponents. In order for the multiplication of two SOPOT-represented scalars having the same amount of SPT terms as its predecessors, the $M$ highest exponents resulting from equation (\ref{eq:multiply}) are retained.

As one observes from (\ref{eq:RLS_error}) to (\ref{eq:RLS_w}), the RLS algorithm requires computing inner products between matrices and vectors. Let ${\bf v}' = [v(0)', v(1)', \dots, v(N-1)']^T$ and ${\bf u}'=[u(0)', u(1)', \dots, u(N-1)']^T$ be two SOPOT-represented line vectors of length $N$, then ${\bf v}'^T{\bf u}'$ accumulates $N$ sums as the one in equation (\ref{eq:multiply}). The product of $N$$\times$$N$ matrix by an $N$-length vector and the product between two $N$$\times$$N$ matrices are implemented using $N$ and $N^2$ inner products.

\vspace*{-.1cm}
\section{Implementation Aspects} \label{sec:implementing}

Let us now describe with further detail how to implement the algorithm in (\ref{eq:RLS_error}) to (\ref{eq:RLS_w}) using  SOPOT arithmetic.

\vspace*{-.1cm}
\subsection{Complexity Upper Bound Constraint}\label{subsec:comp_up_bound_cont}

One can limit the size $B$ of the exponent wordlength and the quantity $M$ of SPT terms. These restrictions are key-elements not only for the complexity reduction but also for the representation precision, thus it affects the quantization error. For example, $B$ will limit the exponent range to be within $L+1-2^B$ and $L$ and therefore the smallest precision is $L+1-2^B$. But this is only achieved if the least significant SPT assumes that value, which in general does not occur.

In each operation the least significant SPT terms are ignored keeping only the $M$ largest ones. This roundoff maintains the complexity restraint but may produce numerical errors.

\subsection{Denominator Inversion}

One may notice in Equation (\ref{eq:RLS_SD}) that two division operations must be performed in the RLS algorithm per each iteration. There is no trivial solution for implementing the division of the denominator of the fraction in SOPOT format. We solve this problem by using approximations for the division operation.

As the denominators in Equation (\ref{eq:RLS_SD}) are scalar values, a division can be interpreted as a multiplication by the inverse of the numbers, i.e., for two scalars $a$ and $b$ we have that $a/b = a.b^{-1}$.
Thus, it is sufficient to approximate the inversion of the denominator. Here, two different methods are proposed to solve this inversion problem. Maintaining the previous notation in which $b$ is the denominator of a fraction, the first case is where $|b| < 2$ and $|b| \geq 2$ is the second case.

\subsubsection{$|b| < 2$}
In this first solution a Maclaurin Series approximation is used, the Maclaurin Series states that for a given scalar $\alpha$, if $| \alpha | < 1$ then:
\begin{align}
    (1-\alpha)^{-1} = \sum_{l=0}^{\infty} \alpha^l. \label{eq:maclaurin}
\end{align}

Applying the Maclaurin Series in a given denominator $b$, represented in SOPOT, one can easily infer that $b = 1 - \alpha \iff \alpha = 2^0 - b$. Even if the series is infinite, the use of a restrained quantity of terms leads to a reasonable approximation of the solution, limiting the sum and implementing the SOPOT notation, Equation (\ref{eq:maclaurin}) can be rewritten as:
\begin{align}
b^{-1} \approx \sum_{l=0}^L (2^0 - b)^l    
\end{align}
where $L \in \mathbb{N}$ and defines the maximum number of terms of the series in the approximation. The greater the absolute value of $b$, the better the approximation with limited $L$.

\subsubsection{$|b| \geq 2$}
Although the first solution is good for the inversion problem, it has the Maclaurin Series' limitation. A second solution is needed when the denominator is outside the Series' validity interval. Thus, for this second solution the most significant SPT term is the only term considered in the inversion. This simplifies inversion since it theoretically does not consist of a SOPOT inversion, but a SPT term inversion. Therefore, the inversion of a scalar $b = 2^{i} \pm 2^{-j} \pm \dots \pm 2^{-k}$, with $i> j > k$. would result in $b^{-1} \approx 2^{-i}$. It is evident that with a significant gap between the two most significant SPT terms a better approximation is obtained.

Both approximation methods for the division operation ensure  that the SPT-RLS algorithm uses only SOPOTs.

\subsection{SOPOT Approximations by Means of the MPGBP}

One must represent the input signal appropriately for complexity reduction using SOPOT filter~\cite{li1994multiplierless,chen1996simplified,ameer2005multiplier}. We impose SOPOT approximation in the tapped delay line ${\bf x}(k)$ vector using the \textit{Matching Pursuits with a Generalized Bit-plane} (MPGBP) algorithm \cite{da2014fir}, as if designing multiplierless FIR filter approximations. The MPGBP approximates an $N$-length vector ${\bf v}$ iteratively as
\begin{align}
{\bf v}' \approx \sum_{i} 2^{\beta_{i}} {\textbf c}_{i}. \label{eq:mpgbp_1}
\end{align}
where $\beta_i$ is the exponent of the SPT term, and ${\bf c}_i$ are codewords whose coordinates are $\pm 1 \textrm{ and } 0$. In \cite{da2014fir} it was shown that for using the least SPT terms, the number of non-null coordinates $P$ per codeword must be $\sqrt{N}$.

\section{Numerical Aspect for the SPT-RLS} \label{sec:nmerical}

In this section, we analyze some impacts of the numerical restrictions and approximations for the adaptive filter.

\subsection{Limited-precision Effects on the RLS Algorithm}

Quantized RLS adaptive filter implementations are possible if the coefficient's quantization error has a zero mean~\cite{ardalan1986floating,cioffi1987limited}. While that result is trivial to derive for some quantization schemes, the previous methods require some analyses. For a random vector $\bf v$, the expected value of its quatization error is
\begin{align}
    E[{\bf e}_q] = E[{\bf v}] - E[{\bf v}']. \label{eq:mean_quant_error}
\end{align}
Let the mean of each coordinate of ${\bf v}$ be zero. Since $\bf v$ is approximated using the MPGBP, one has that
\begin{align}
    E[{\bf e}_q] = - E\left[ \sum_i 2^{\beta_i} {\bf c}_i \right]
\end{align}
where each ${\bf c}_i$ has $\sqrt{N}$ non-null coordinates (1 and -1). If the probabilities of the coordinates being 1 or -1 independently of $\beta_i$, then the quantization error for every coordinate is zero.


In order to validate the previous rationale and conclusion, a numerical analysis of the quantization error is proposed.
For this analysis, the MPGBP algorithm approximates random vectors of different lengths $N$, the process was repeated $10^5$ times in a Monte Carlo method. The maximum quantization error of a vector coefficient for some combinations of $M$ and $N$ are presented in Table \ref{tab:expect_val}. It is observable that for the smallest $M/N$ the mean of the quantization error is the largest but it is already low (of order $10^{-5} \approx 2^{-17}$).



\begin{table}[ht]
\caption{Expected value and standard deviation of quantization error of SOPOT approximations by means of the MPGBP for some restrictions on the total amount of SPT terms $M$ per vector element for different vector lengths $N$.}
\label{tab:expect_val}
\centering
\begin{tiny}
\begin{tabular}{|c|c|c|c|c|}
\hline
$(M/N)$ ~~ \textbackslash{} ~~ $N$ & $4$            & $12$          & $16$          \\ \hline
$3$ &\begin{tabular}[c]{@{}l@{}}$-1.99 \times 10^{-5}$\\ $\pm 7.02 \times 10^{-5}$\end{tabular} & \begin{tabular}[c]{@{}l@{}}$1.29 \times 10^{-5}$\\ $\pm 1.22 \times 10^{-5}$\end{tabular} & \begin{tabular}[c]{@{}l@{}}$-2.06 \times 10^{-5}$\\ $\pm 1.03 \times 10^{-5}$\end{tabular} \\ \hline
$5$ &\begin{tabular}[c]{@{}l@{}}$1.02 \times 10^{-6}$\\ $\pm 3.48 \times 10^{-8}$\end{tabular} & \begin{tabular}[c]{@{}l@{}}$-3.33 \times 10^{-7}$\\ $\pm 4.09 \times 10^{-9}$\end{tabular} & \begin{tabular}[c]{@{}l@{}}$3.74 \times 10^{-7}$\\ $\pm 3.27 \times 10^{-9}$\end{tabular} \\ \hline
$7$ &\begin{tabular}[c]{@{}l@{}}$1.58 \times 10^{-8}$\\ $\pm 1.74 \times 10^{-11}$\end{tabular} & \begin{tabular}[c]{@{}l@{}}$6.92 \times 10^{-9}$\\ $\pm 1.30 \times 10^{-12}$\end{tabular} & \begin{tabular}[c]{@{}l@{}}$-9.10 \times 10^{-9}$\\ $\pm 9.78 \times 10^{-13}$\end{tabular} \\ \hline
$10$ &\begin{tabular}[c]{@{}l@{}}$-8.78 \times 10^{-11}$\\ $\pm 1.89 \times 10^{-16}$\end{tabular} & \begin{tabular}[c]{@{}l@{}}$-2.29 \times 10^{-11}$\\ $\pm 7.78 \times 10^{-18}$\end{tabular} & \begin{tabular}[c]{@{}l@{}}$-1.38 \times 10^{-11}$\\ $\pm 5.24 \times 10^{-18}$\end{tabular} \\ \hline
\end{tabular}
\end{tiny}
\end{table}

\begin{figure}
    \centering
    \includegraphics[width=\linewidth]{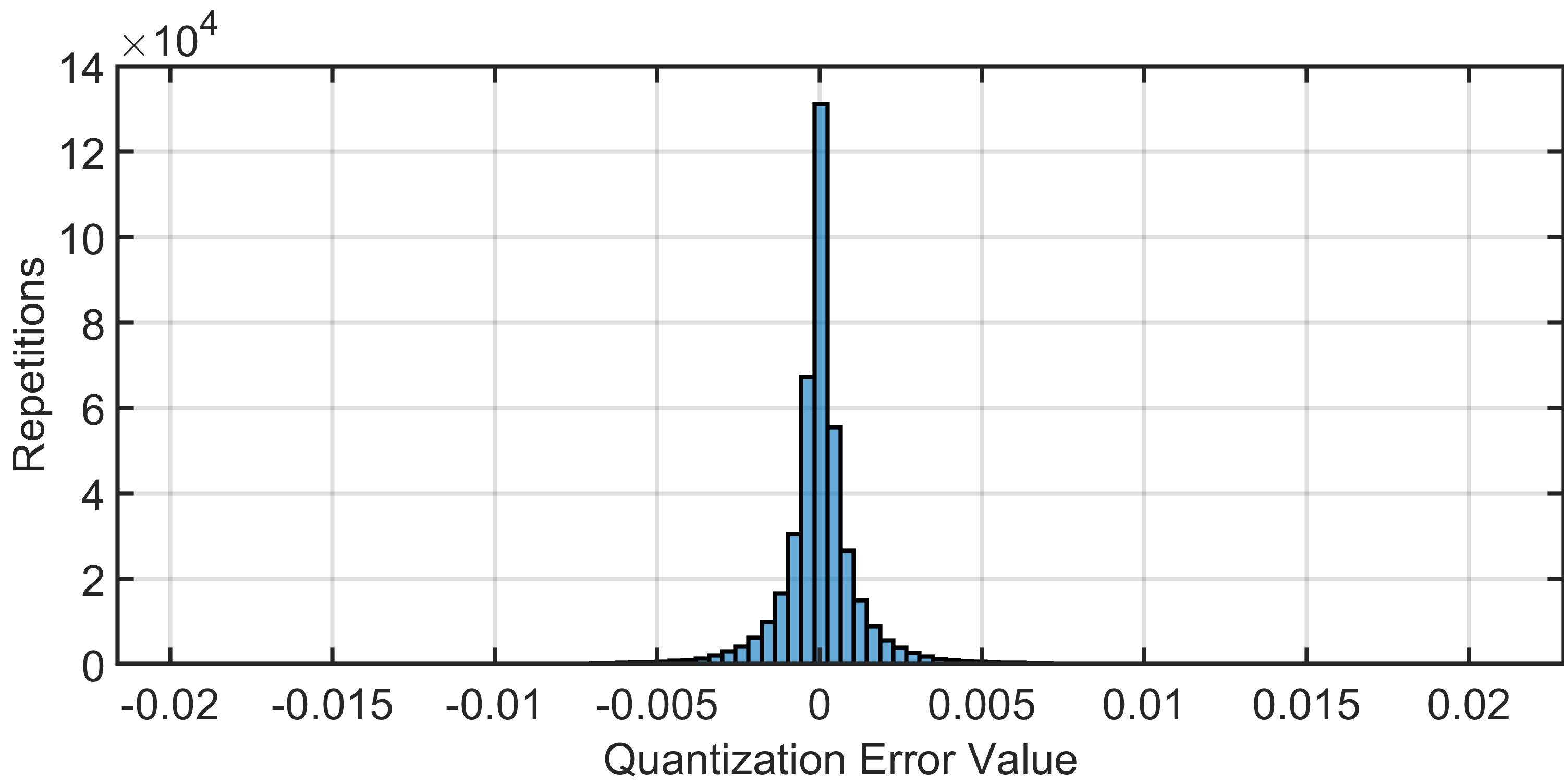}
    \caption{Histogram (using $100$ bins) of SOPOT approximations by means of the MPGBP considering all the elements of a $N$=4 vector with $M/N$=4.}
    \label{fig:hist}
\end{figure}

Fig. \ref{fig:hist} presents a 100-bin histogram containing the distribution of the obtained quantization error for a vector with $N$=4 using the MPGBP with $M/N$=4. One can observe the high distribution of values near zero.

\subsection{Complexity Upper Bound Constraint Impacts}

In order for the adaptive filter to converge, it is also important that the quantization error remains with a zero mean throughout the iterations. To guarantee complexity reduction, the result of each numerical operation is roundoff as discussed in Section~\ref{subsec:comp_up_bound_cont}.
A priori, there is no reason to imagine that the probabilities of truncating positive and negative SPTs are not the same. This rationale is supported by the values that SPTs may assume. Thus, the complexity restriction is expected not to hinder the zero mean roundoff error assumption. 

\section{Results} \label{sec:results}

The SPT-RLS adaptive filter is tested in a system identification scenario and its results are compared with those of a 16-bit fixed-point RLS adaptive filter using 11 fractional bits.

In the simulation the input signal $x(k)$ is Gaussian white noise, the desired signal $d(k)$ is $x(k)$ convolved with the unknown system $H$ and undergoes Gaussian white noise $n(k)$, here $H$ is a pseudo-randomly generated 4-element vector. We expect the adaptive filter to mimic $H$'s elements. Although, the additive noise $n(k)$ is uncorrelated with $d(k)$ thus a version of this signal is expected to be present in the error signal $e(k)$.

For $x(k)$ with $K$=300 samples, the adaptive filter has $N$=4 coefficients, for the fixed-point RLS the forgetting factor $\lambda$=0.9987 and for the SPT-RLS $\lambda$=0.9533, the initialization value for the deterministic correlation matrix of both cases is $\delta$=0.09.
The SPT-RLS exponent wordleght $B$=4. The additive noise $n(k)$ was adjusted for SNR=3 and 20 dB, representing different working conditions. The simulation was repeated 100 times for more reliable results.

\begin{figure}[hbt!]
    \centering
    \includegraphics[width=1\linewidth]{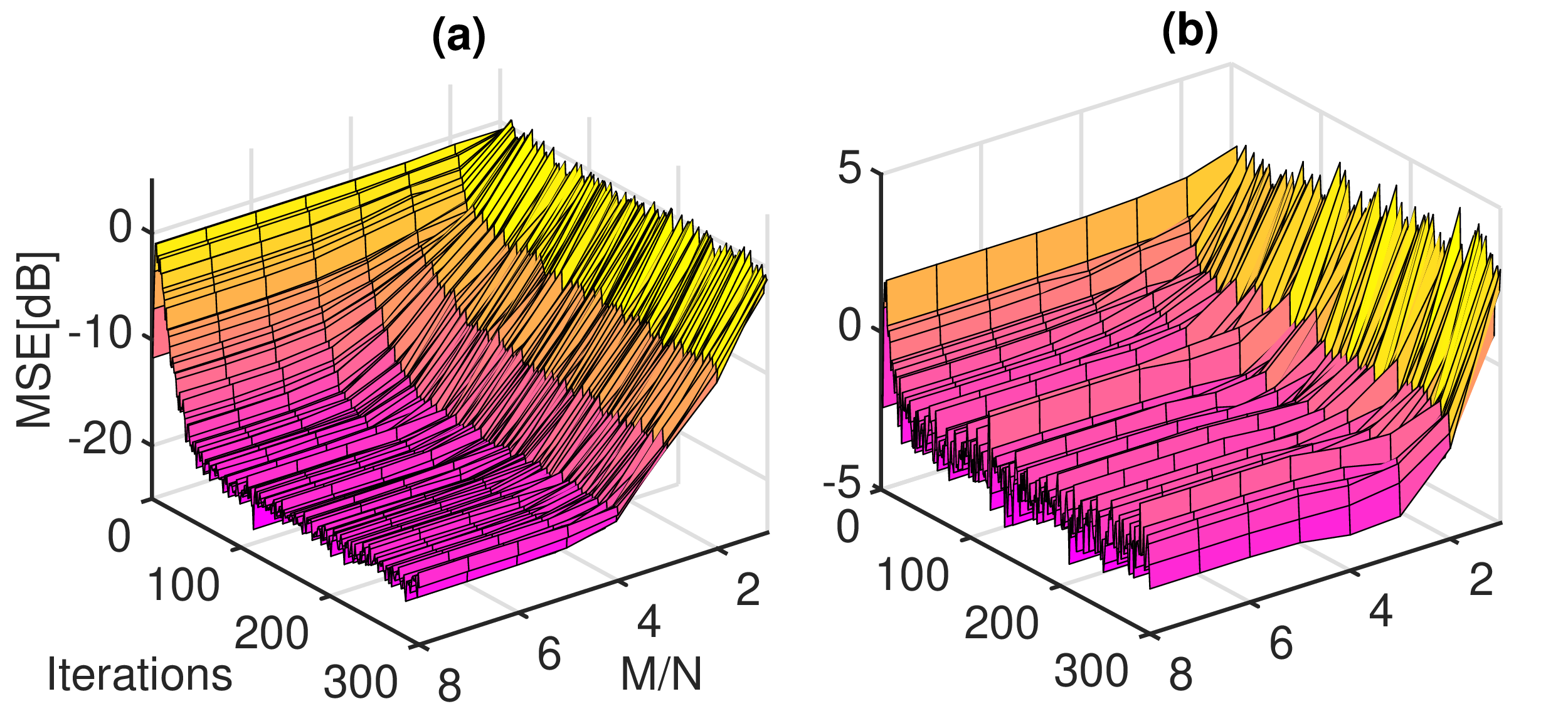}
    \caption{MSE evolution surface for the SPT-RLS adaptive filter with different complexities, where in $\bf (a)$ SNR=20 and $\bf (b)$ SNR=3.}
    \label{fig:MSE_SPT-RLS}
\end{figure}

The MSE curve for each different amount of SPT terms per coefficient $M/N$ is evaluated in the surface of Fig.~\ref{fig:MSE_SPT-RLS}. This can be used to evaluate the influence that the amount of $M$ SPT terms per coefficient has in the SPT-RLS algorithm. If $M/N$ is too restrictive then the algorithm will not be able to converge, one notes that after a certain complexity $M/N$ is achieved, the filter starts to present the desirable results.

\begin{figure}[hbt!]
    \centering
    \includegraphics[width=1\linewidth]{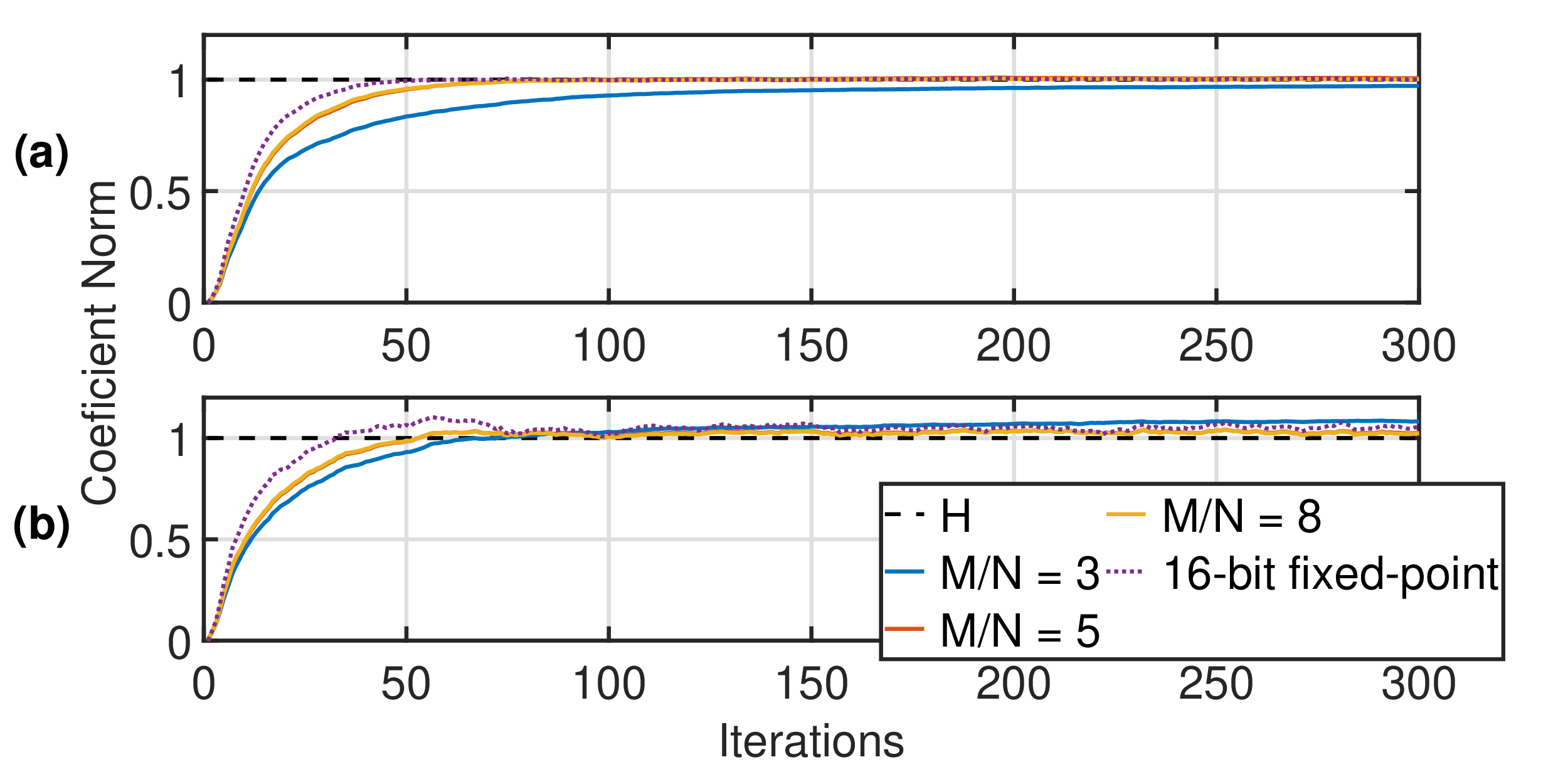}
    \caption{The coefficients norm for different complexities levels of the SPT-RLS algorithm throughout the iterations, where in $\bf (a)$ SNR=20 and $\bf (b)$ SNR=3.}
    \label{fig:coeffNorm}
\end{figure}

\begin{figure}[hbt!]
    \centering
    \includegraphics[width=1\linewidth]{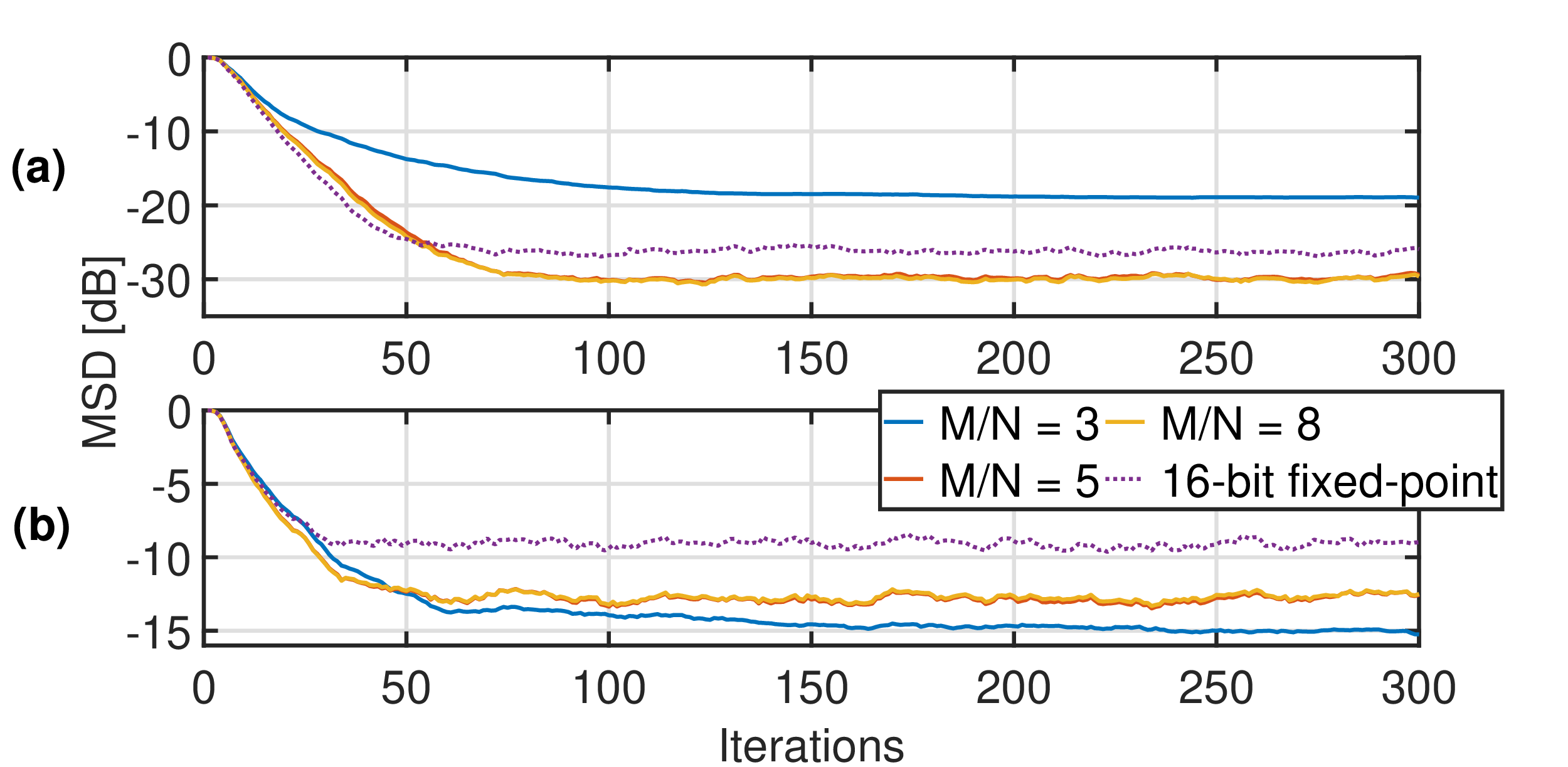}
    \caption{MSD between the filter $H$ and the SPT-RLS adaptive filter for $M/N$=3, 4 and 5, also between the filter $H$ and the 16-bit fixed-point RLS adaptive filter, with the fraction length of 11 bits, where in \textbf{(a)} SNR=20 and \textbf{(b)} SNR=3.} 
    \label{fig:deviation}
\end{figure}

To evaluate the adequacy of the SPT-RLS filters one compares them to the their fixed-point counterparts. As the filters try to mimic the filter $H$, the coefficients vector norm as the filter adapts is presented in Fig.~\ref{fig:coeffNorm}. As expected, it is noticeable that the SPT-RLS adaptive filter closely mimicked the system, and provided results that are very similar to the ones obtained with the 16-bit fixed-point RLS. In Fig.~\ref{fig:coeffNorm} one cannot clearly evaluate how well $H$ is approximated by the other filters, although it can be properly evaluated with the Mean Squared Deviation (MSD) parameter, which evaluates the difference between the coefficients and $H$. In Fig.~\ref{fig:deviation} the MSD for the RLS-SPT with different complexities and for the precision-equivalent (16-bit) fixed-point RLS are presented.

For the complexity evaluation of the fixed-point implementation only the multiplications were evaluated, and the considerations were generous regarding its complexity. Let the fixed-point symbol-length be $B_F$ bits. The multiplication between two symbols requires that one of the symbols is shifted in function of the non-null bits of the other and the addition of the resulting shifted binary words, that may occupy $2B_F$ bits. Assuming a sufficiently long symbol and that the encoded values are distributed symmetrically over the fixed-point range, then the average number of active bits per fixed-point symbol is $B_F/2$. This reduces the complexity of the multiplication between two fixed-point symbols to be $B_F/2$ shifts of $B_F$-bit words and $B_F/2-1$ additions of possibly $2B_F$-bit words. Using $M$ SPT terms and $B$-bit exponents, the same multiplication operation requires at most $M^2$ additions (distributive property) of $B$-bit words. Regardless of storage requirements, one may compare the complexities of both representations for the multiplication of two numbers using the amount of necessary ``1-bit sums''. Thus, for the fixed-point $B_F^2-2B_F$ sums are required per multiplication and for the SOPOT $BM^2$ sums. Therefore, to evaluate performances under similar computational complexities, we compare the SPT-RLS with the fixed-point implementation using $B_F = 1 + \sqrt{1+BM^2}  \approx \left \lceil 1 + M\sqrt{B} \right \rceil $ bits.     

One can notice in Fig.~\ref{fig:deviation} that, for a same estimated precision, the results for the SPT-RLS are slightly better. For the scenario leading to Fig.~\ref{fig:deviation}.\textbf{(a)} there is no gain in using more than $M/N$=5 SPT terms, indicating a possible limit on the required complexity for the SPT-RLS adaptive filter.

\begin{figure}
    \centering
    \includegraphics[width=1\linewidth]{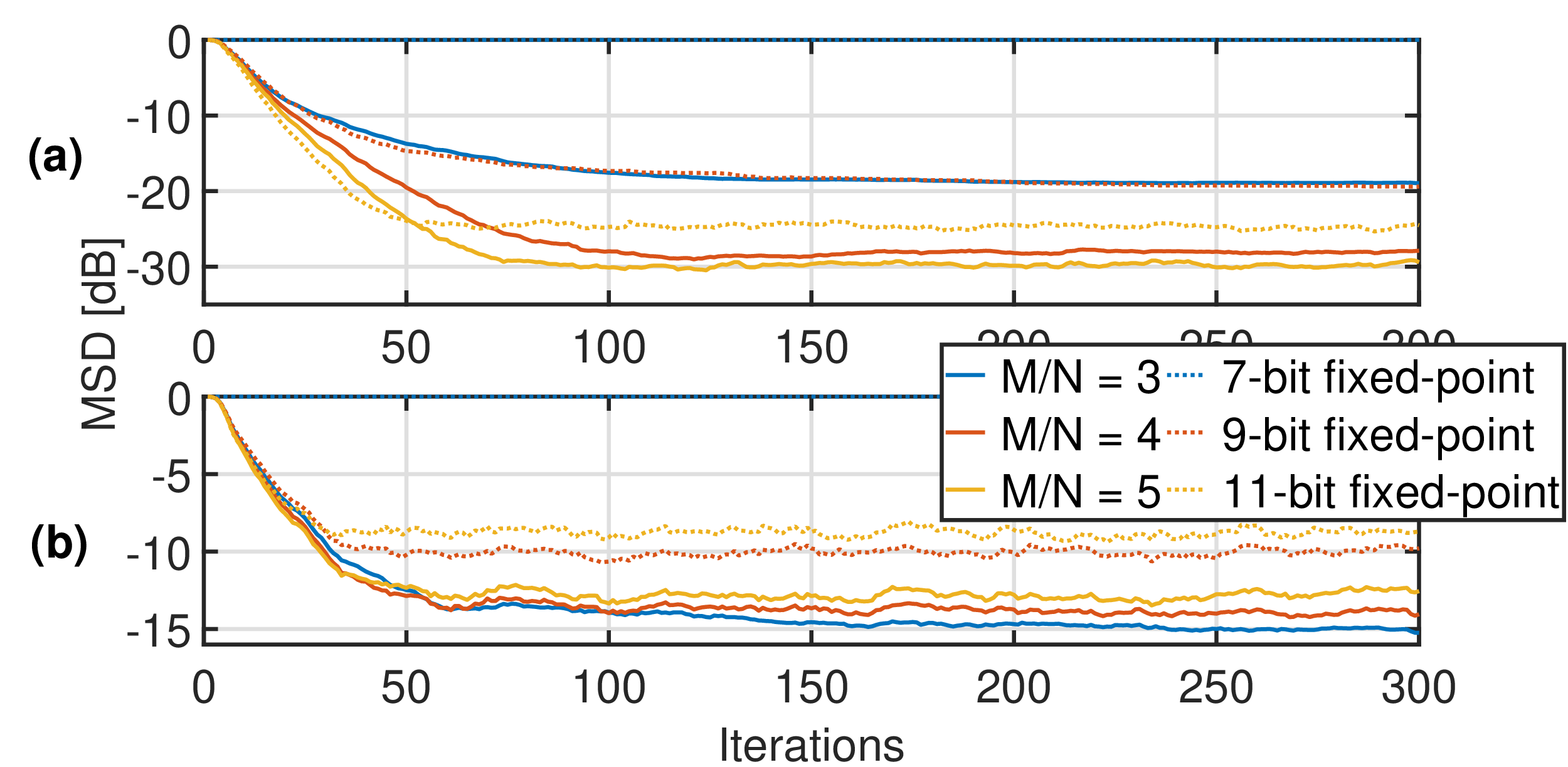}
    \caption{MSD between the filter $H$ and the SPT-RLS adaptive filter for $M/N$=3, 4 and 5, also between the filter $H$ and the 7-bit, 9-bit and 11-bit fixed-point RLS with 2-bit, 4-bit and 6-bit fraction lengths respectively, where in \textbf{(a)} SNR=20 and \textbf{(b)} SNR=3.}
    \label{fig:dev_complex}
\end{figure}

In Fig.~\ref{fig:dev_complex}.\textbf{(a)}, one observes that the convergence times vary with the amount of SPT terms and that also different post-convergence results are obtained. In addition, one may compare the fixed-point- and the SPT-RLS filters. The 7-bit fixed-point was not capable of converging although its SPT-RLS counterpart ($M/N$=3) achieved approximately the same result that the more complex 9-bit fixed-point did. The SPT-RLS obtains better results when compared with the fixed-point RLS of similar computational complexity. From the different SNR levels analyses one can infer that a less privileged scenario may require fewer SPT terms from the SPT-RLS.

\section{Conclusion} \label{sec:conclusion}

The proposed SPT-RLS is a functional multiplierless adaption application capable of providing adaptive filters with reduced computational burden both in the algorithm and in the resulting filter. The obtained results show that the nonuniform quantization obtained through the SOPOT approximation does not make the adaptive filter impracticable. Furthermore, the proposed fixed-point like structure for the SPT-RLS is capable of obtaining distinctly superior results than the fixed-point RLS counterpart with significantly reduced complexity.

\clearpage

\bibliography{biblio.bib}
\bibliographystyle{IEEEtran}

\end{document}